\documentstyle[12pt]{article}
\begin{document}


\title{Local and Global Gravity\footnote
{UCONN 96-09, November 1996}}

\author{\normalsize{Philip D. Mannheim} \\
\normalsize{Department of Physics,
University of Connecticut, Storrs, CT 06269} \\ 
\normalsize{mannheim@uconnvm.uconn.edu} \\
\normalsize{} \\ 
\normalsize{To appear in a special issue of Foundations of Physics} \\ 
\normalsize{honoring Professor Lawrence Horwitz on the occasion of his 
65th birthday;} \\ 
\normalsize{A. van der Merwe and S. Raby, Editors, Plenum 
Publishing Company, N.Y., 1997.} \\}

\date{}

\maketitle

\begin{abstract}
Our long experience with Newtonian potentials has inured us to the 
view that gravity only produces local effects. In this paper we 
challenge this quite deeply ingrained notion and explicitly identify 
some intrinsically global gravitational effects. In particular we 
show that the global cosmological Hubble flow can actually modify the 
motions of stars and gas within individual galaxies, and even do so 
in a way which can apparently eliminate the need for galactic 
dark matter. Also  we show that a classical light wave acquires an 
observable, global, path dependent phase in traversing a 
gravitational field. Both of these effects serve to underscore the 
intrinsic difference between non-relativistic and relativistic 
gravity. 
\end{abstract}

\section{Introduction}

Since its very inception gravitational theory has always been regarded as being 
intrinsically local. Specifically, with the Newtonian gravitational potential 
falling as $1/r$, the motion of any given system is then primarily determined by 
the most nearby gravitational sources, with static, spherically symmetrically 
distributed distant matter actually making no net contribution at all. Thus in a 
static, spherically symmetric Newtonian universe only nearby sources are 
relevant. Moreover, the advent of Einstein's general theory of relativity 
actually only served to reinforce this viewpoint, since the metric inside a 
static (or even a radially vibrating) spherically symmetric shell is found to be 
flat in Einstein gravity, to thus again permit us to locally ignore the rest of a 
universe of the same symmetry. Thus firstly with potentials which fall to zero at 
large distances and then subsequently with the analogous relativistic notion of 
asymptotic flatness, a view of gravity has emerged that only local sources are 
relevant. In this paper we examine and challenge this longstanding and by now 
quite ingrained notion by identifying some explicitly global aspects of gravity, 
both in standard Einstein gravity itself, as well as in the recently advanced 
fourth order conformal invariant alternate theory of gravity (see e.g. 
\cite{Mannheim1994}), a theory in which individual gravitational sources are 
found to actually make asymptotically non-vanishing contributions to the geometry, 
and in which the global cosmological Hubble flow actually has local consequences, 
to thus give gravity a quite Machian flavor.

Since a great deal of our gravitational intuition comes from Newtonian celestial 
mechanics, we begin in Sec. (2) with an analysis of the observational basis
for inferring that non-relativistic potentials actually fall at large distances;
and show that rising ones are not merely not excluded by available data,
but that a phenomenological analysis of the systematics of galactic rotation 
curves indicates that (linearly) rising potentials may in fact even 
be favored. Moreover, our specific study reveals that not one but in 
fact two linear potentials appear to be operative in galaxies, one which scales 
linearly with the amount of matter in a given galaxy, and a second one which   
intriguingly has a universal, galaxy independent normalization of order the
Hubble scale. In Sec. (3) we show that relativistic covariance does not in fact 
lead us unambiguously to standard Einstein gravity, and that it is possible to 
consider other, equally covariant, pure metric theories of gravity, theories 
which could then have totally different asymptotic properties. In particular we 
show that the two phenomenologically found linearly rising galactic 
potentials of Sec. (2) even have a natural home in the conformal gravity 
alternative where they can both be produced, one locally by the luminous matter 
within a given galaxy and the other globally by the universal Hubble 
flow due to the rest of the galaxies in the universe. In Sec. (4) we show that 
the COW (Colella, Overhauser and Werner) quantum-mechanical neutron beam 
interferometry experiment in a background gravitational field has a purely 
classical counterpart, with classical light waves also experiencing global, path 
dependent interference phase shifts in background gravity, to thus yield another 
global, expressly relativistic effect; and with the interference being found to 
be accreditable to the gravitational bending of light, this effect is then one 
which will occur in any covariant, pure metric based theory of gravity. From 
the various specific studies of this paper there thus emerges a global 
character to gravity of a surprisingly far reaching nature.  

\section{Local and global non-relativistic gravity}

In Newtonian gravity the $-MG/r$ potential has four very remarkable properties.
Firstly, it permits the replacement of a static, spherically symmetric 
source of non-zero radius by a point source located at its center. Secondly, it 
provides for the mutual cancellation of the potentials of distant, static, 
spherically symmetrically distributed matter sources, so that such matter has no 
net local effect at all. Thirdly, since $-1/4\pi r$ is the Green's function of 
the $\nabla ^2$ operator, the $-MG/r$ potential also emerges as the exterior 
solution to the second order Poisson equation, viz.
\begin{equation}
\nabla^2 \phi(r)=4\pi G\rho(r)~~~,
~~~\phi(r>R_0)=-{4\pi G \over r}\int_0^{R_0}dr^{\prime}
\rho(r^{\prime}){r^{\prime}}^2
\label{(1)}
\end{equation}
for a static, spherically symmetric source of radius $R_0$ and mass density 
$\rho(r)$, to thus connect the Newtonian potential to a dynamical equation of 
motion; and uniquely so in fact within the framework of second order equations,
since no other power behaved function $r^{-n}$ is simultaneously both the input 
Green's function and the output potential to a second order differential 
equation of motion except $n=1$. And fourthly, of course, the $-MG/r$ potential 
provides an excellent first approximation to the celestial mechanics of the solar 
system as well as to all non-relativistic gravitational phenomena on all smaller 
distance scales (such as terrestrial), with its only real shortcoming in this 
regard being the phenomenological existence of the (small) planetary orbit 
precessions which a $1/r$ potential strictly forbids. Now even though all four of
these features have been instrumental to our thinking and to our accepting of 
Newtonian gravity, in a sense the first three features are somewhat moot with 
only the fourth one actually bearing directly on observation. Thus, for instance, 
while it is very nice to be able to replace the sun by a point source, nonetheless 
this is only a calculational convenience. Moreover, even while a static, 
spherically symmetric universe may indeed decouple locally, nonetheless we do not 
actually live in such a universe, with there currently being no such analogous 
decoupling theorems for other possible cosmologies such as comoving, expanding, 
globally topologically non-trivial ones. And finally, 
even if we accept the whole four features as representing basic properties of 
gravitational theory, such Newtonian considerations then lead us to the notion 
that gravitational sources are to have negligible effect at infinity, a 
conclusion which even though it is widely accepted has nonetheless never in fact 
actually been tested, with it not being at all clear as to how it might  
actually even be tested since the universe in which we live is not in fact 
asymptotically flat. The locality of gravity is for the moment then essentially a 
preference of the community rather than an established fact.

Nonetheless, all of this Newtonian wisdom notwithstanding, as of today the issue 
of whether or not this same $1/r$ potential law actually continues to fit data on 
distance scales much larger than that of the solar system is actually still a live 
and yet to be resolved one. Specifically, it is found that on galactic distance 
scales the velocities of galactic stars and hydrogen gas (and particularly the gas 
since it is generally distributed in galaxies out to much farther distances from  
galactic centers than the stars) do not show any sign at all of the Keplerian
fall off associated with a $1/r$ potential. Indeed, as can be seen from Fig. (1) 
which shows a typical set of galactic rotation curves, at the largest detected 
distances (distances which are well beyond the detected galactic luminous matter 
regions) the observed orbital rotational velocities are way above the falling 
luminous Newtonian expectation, and are themselves not only far from falling, but 
in fact are generally seen to be close to flat or in some cases to even be rising. 
Now while this shortfall in the velocities has prompted the notion of galactic 
dark matter (viz. something purely local as per the standard Newtonian reasoning) 
which is to then provide some additional $1/r$ potential contributions which are 
to bring the rotation curve predictions back into line with data, nonetheless 
this appeal to dark matter is only makable provided it is a priori known that 
Newton's law of gravity actually continues to hold on these very big distance 
scales. However, as of today, there has not yet been any independent 
confirmation of Newton's law of gravity on these distance scales whatsoever, 
i.e. one which only involves expressly falsifiable predictions based on 
unambiguously established 
matter sources with observationally established density distributions, since 
galactic dark matter always has to be invoked, to thus show the complete 
circularity of the reasoning.\footnote{Moreover, even if one accepts dark 
matter as a general concept, in reality one also has to make a guess as to its 
specific spatial distribution with dark matter fitting of the 11 galaxies shown 
in  Fig. (1) entailing the need for 2 independent parameters per dark matter 
halo and thus 22 in all for the whole set, a far from satisfactory situation.} 
In fact, if galactic rotation curves were the only data available to us (i.e. 
if we had no solar system information at all) we would not be able to extract 
out a $1/r$ potential law at all. The continued use of Newton's law of gravity 
on galactic distance scales for the moment thus rests on a not yet completely 
secure foundation. However, if one is actually prepared
to contemplate trying to go beyond Newtonian gravity (and therefore by 
implication to go beyond its covariant Einstein generalization, a point we 
address in detail below), then not only does one have to deal with the difficult 
issue of actually challenging the standard theory at all, the only way to 
modify Newton galactically would be to have to have some new effect which would 
then dominate over Newton at these large distances, thus making the Newtonian 
term no longer dominant at infinity and thereby opening up the possibility that 
gravity might then have global effects. Despite the fact that this possibility 
goes completely against our local Newtonian intuition, we 
shall now show that it actually has some observational support.

While a first glance at Fig. (1) reveals some strikingly flat rotation 
curves,\footnote{This set of galaxies has been identified by Begeman, Broeils 
and Sanders \cite{BegemanBroeilsandSanders1991} as possessing particularly 
reliable data which exhibit the basic pattern of departure from the luminous 
Newtonian expectation that has so far been observationally obtained, with their 
paper giving complete data references.} it is more instructive for fundamental 
theory to look not at the total velocities, but rather at the velocity 
discrepancies, viz. the excess of the total velocity over the luminous Newtonian 
contribution. As we see, these discrepancies themselves are far from flat, and 
in fact are actually growing with distance $R$ from the center of each galaxy. Now 
while this must be the case for galaxies whose total velocities are flat (since 
the discrepancy is the excess over the falling luminous Newtonian contribution),
what makes this remark so significant is that it also applies to the low
luminosity galaxies as well (the galaxies are presented in order of increasing 
luminosity in Fig. (1)), galaxies whose total velocities are currently found
to predominantly be rising, and to not in fact be flat.\footnote{While none of the 
low luminosity galaxies currently show any flat rotation curve region at all, 
there is a noticeable turnover in one of these galaxies, viz. DDO 154, the 
galaxy which is altogether the least luminous of the whole sample. However, 
since this is the most gas dominated galaxy in the sample, random 
gas pressures could be making a substantial contribution to motions in 
the turnover region. We shall thus ignore any possible ramifications of 
these last few points here, though clearly if this turnover proves to be 
a real trend which is then reproduced in other low luminosity galaxies, 
it would eventually have to be accounted for.} The fact that the 
discrepancies are rising out to the last detected points in each and every 
galaxy in the set (a set whose luminosities vary by a factor of more than 
1000 - see Table (1) which lists some specific input data) thus emerges as a 
universal feature of the data even as the total velocities as a set do not show 
any common universal underlying trend.\footnote{The flatness of the total 
velocities of the higher luminosity galaxies is so striking that the rise in 
the lower luminosity galaxies has generally been discounted by the community as 
being in any way representative of a possible trend. And even though the ratios of  
the velocity discrepancies to the luminous Newtonian contributions are actually 
greater in the low luminosity galaxies than in the higher luminosity ones (thus 
incidentally forcing dark matter model fits to require that the dark to luminous 
mass ratio be bigger for the lower luminosity galaxies - an assumption of dark 
matter theory which is still not all that well understood), 
nonetheless it is still taken as a 
given that the total velocities in the low luminosity galaxies will eventually 
flatten off. However, the fact that the discrepancies are currently rising in all 
the galaxies suggests a possibly different asymptotic outcome, with there 
currently being no observational support at all for the widely held belief that 
the discrepancies (in both low and high luminosity galaxies) will in fact all be 
flat asymptotically.}

In order to universally quantify this qualitative velocity discrepancy trend, 
it is very instructive \cite{Mannheim1995a}, \cite{Mannheim1996a} to evaluate the 
total centripetal acceleration 
$(v^2/c^2R)_{last}$ at the last data point in each galaxy (except for DDO 154 
for which we use the last point before the turnover). As we can see 
from the fourth column in Table (1)  $(v^2/c^2R)_{last}$ (a completely model 
independent quantity) turns out to be remarkably universal, varying 
only by a factor of 5 or so over the sample, a variation which is altogether 
less than the factor of 1000 or so by which the luminosity varies in the same 
sample, so that $(v^2/c^2R)_{last}$  emerges as a quantity which contains
universal information. Moreover, we see a small but clear trend with 
increasing mass in the centripetal accelerations. And in fact we find 
\cite{Mannheim1996a} that we can parameterize all the 11 total accelerations
listed in Table (1) according to the one universal three component relation 
\begin{equation}
(v^2/c^2R)_{last}=\gamma_0/2+\gamma^{*}N^{*}/2 
+\beta^{*}N^{*}/R^2 
\label{(2)}
\end{equation}
where the two new universal constants $\gamma_0$ and $\gamma^{*}$ take numerical 
values $3.06\times 10^{-30}$ cm$^{-1}$ (a value strikingly close to the inverse 
of the Hubble radius) and $5.42\times 10^{-41}$ cm$^{-1}$ respectively, where
$\beta^{*}=1.48\times 10^5$ cm, and where $N^{*}$ is the total amount 
of visible stellar (and gaseous) material in solar mass units in each 
galaxy. Before venturing to discuss the possible significance of this new, 
purely phenomenological, model independent regularity, it is important to 
realize that there is nothing in any way special about the actual magnitudes 
of the radial coordinates, $R$, of the last detected points in the 11 galaxies, 
since their locations are fixed purely by the instrumental limits of the 
various detectors used in measuring the various gas surface brightnesses and 
not fixed by any dynamics associated with the galaxies themselves. Thus the 
magnitude of each last measured radial $R$ (a quantity which varies from 8 
kpc to 40 kpc or so over the sample) is essentially arbitrary for the 
galaxies, and yet $v^2/R$ can nonetheless still be universally parameterized. 
As far as we can see, with the luminous Newtonian contribution being 
decidedly non-leading at the farthest data points, the only obvious way that 
this could in fact occur would be if $v^2$ were in fact growing universally 
with $R$ so that the magnitude of $v^2/R$ would not in fact depend on where 
the last detected points just happened to be located within galaxies. This 
pattern is clearly not one that one would expect with flat rotation curves, or 
even in fact think to look for in such a paradigm, and would instead seem to 
point to potentials which if anything are actually growing (linearly) with 
distance rather than falling in the familiar Newtonian manner. Moreover, if we 
provisionally identify $\gamma^{*} c^2r/2$ as the linear potential put out by a 
typical star such as the sun (our conformal gravity study below will actually 
justify this identification), then the numerical value for $\gamma^{*}$ as 
inferred from Eq. (\ref{(2)}) is so small that the contribution made by such a 
linear potential to solar system dynamics is then totally negligible, to thus 
yield us phenomenological linear potentials for stars which only start to 
become competitive with their Newtonian ones on galactic distance scales, these 
intriguingly being the distance scales where the standard theory first has to 
appeal to dark matter. Thus we see from purely phenomenological considerations 
alone that it is somewhat unwarranted to yet claim 
the validity of Newton's law of gravity on distance scales much larger than 
the solar system ones on which it was first established, and that despite 
a preconceived prejudice to the contrary, currently available data do not in 
fact exclude potentials which grow rather than fall with distance. In fact,
since the data even appear to lend some support to this possibility, it is 
to the theoretical underpinnings needed for such an option that we now turn,
underpinnings which in fact will enable linear potentials to account for the 
systematics of the velocity discrepancies of all the data points in Fig. (1) 
and not just merely of those of the last ones.   

\section{Local and global relativistic gravity}

Even though we have just made a phenomenological case for rising potentials,
nonetheless, since it was based entirely on non-relativistic reasoning, in
order for it to have any possible fundamental significance it is necessary
to show that it is compatible with relativity. Now as regards relativity
itself, it is important to realize that the validity of the equivalence principle
(the cornerstone of relativistic gravity) is secured once the metric tensor
$g^{\mu \nu}(x)$ is identified as the gravitational field, with no commitment 
being needed as to the specific equations of motions which $g^{\mu \nu}(x)$ is to 
then obey, save only that they be covariant. The principle of general relativity
thus requires only that the gravitational action be a general coordinate scalar
function of the metric, with there being no less than an infinite number of
such possible candidate scalar actions involving derivatives of the metric of 
arbitrarily high order. Motivated by the successes of Newtonian gravity 
Einstein himself opted for a theory based on second order derivatives of the 
metric, viz.
\begin{equation}
R^{\mu \nu} - g^{\mu \nu} R^{\alpha}_{\phantom {\alpha} \alpha}/2
= -8 \pi G T^{\mu \nu}
\label{(3)}
\end{equation}
an equation of motion whose weak gravity limit is the second 
order Poisson equation of Eq. (\ref{(1)}), to thus both yield Newton's law of 
gravity and to provide some relativistic corrections to it, corrections which
then completely account for the problem of the precessions of the planetary 
orbits to which we referred earlier. 

On an even deeper level the metrication of gravity also provided for a coupling 
of light to gravity, and since light travels in waves which fill all space,
it thus gave gravity some global aspects absent in its coupling to material
point particles. Since the global aspects of the coupling of light to gravity 
will prove central to our discussion below of the COW interferometry experiment, it
is of some value to discuss the nature of this coupling from the point of view
of the equivalence principle. Even though the coupling of light to gravity is 
motivated by Einstein's elevator, this is not actually the origin of the 
coupling. Specifically, while one can conclude from an analysis of a freely 
falling observer in an elevator that such an observer could not distinguish
between a uniformly accelerating elevator and a uniform gravitational 
acceleration, that observer is only unable to make such distinction because of
the equality of inertial and gravitational masses. From this fact it does not
necessarily follow that the same observer could not make any distinction for 
light since light is massless, and thus may not couple to gravity at all (i.e.
the equation of motion could degenerate into zero equals zero), with 
the equivalence principle then simply not applying to it, and with light
then bending in an accelerated coordinate system but not in a gravitational
field, an inelegant but nonetheless logical option.\footnote{A specific model 
where this would in fact be the case would be a scalar theory of gravity 
in which the source of the scalar field is the trace of the energy momentum 
tensor, a quantity which simply vanishes for light. The author is indebted to 
Dr. A. Chodos for alerting him to such a model.} As an alternative to starting  
by simply postulating that there is an equivalence principle for both material 
particles and for light, one can instead proceed just as one does in special 
relativity. Thus in the same way that special relativity is based on the 
requirement that all uniformly moving observers have to agree on the same 
physics, then the analogous starting requirement in general relativity would  
be to require that non uniformly moving observers also all have to agree on the 
same physics, to thus lead us to general coordinate invariance (but not yet 
necessarily to curved space since the assumption of coordinate invariance alone 
does not suffice to force the Riemann tensor to actually be non-zero). Moreover, 
in and of itself the principle of coordinate invariance does not tell us how the 
gravitational field itself is to be described, since at this point it could be 
just some arbitrary field defined over spacetime just like the electromagnetic 
field (with gravitational sources being treated analogously to electromagnetic 
ones and with the universe then being covariant but Riemann flat). It is only 
the further, independent assumption that the gravitational field is not any old
field but is instead the very special one associated with the spacetime 
metric\footnote{If no physical field were ever associated with the metric, there 
would then be no equation of motion with which to fix the curvature of 
spacetime at all, be it flat or not. Thus once given the imposition of a principle 
of general covariance it would appear that the metric would then have to be 
dynamical.} that then leads us to a covariantly coupled metric theory of gravity 
(and thus a covariant coupling to electromagnetism as well) in which geodesic 
motion and the equivalence principle are then output to the theory (rather than 
input) for both massive and massless particles, so that light then does indeed 
obey the equivalence principle after all.\footnote{The solution to the planetary 
precession problem could in principle have been found within the framework of 
Newtonian gravity by a small departure from strictly $1/r$ potentials. It was 
only the experimental detection of the gravitational bending of light which 
assured that conceptual departures from Newtonian gravity were unavoidable.} It 
is in this way that gravity thus acquires global features.

As regards the standard classic tests of Einstein gravity, they are all 
nicely contained within the familiar Schwarzschild $R^{\mu \nu}=0$ vacuum 
metric solution     
\begin{equation}
d\tau^2=(1-2MG/c^2r)c^2dt^2-dr^2/(1-2MG/c^2r)-r^2d\Omega
\label{(4)}
\end{equation}
exterior to a static, spherically symmetric gravitational source of mass $M$.
This solution then establishes asymptotic flatness as a 
characteristic feature of standard gravity, to thus dovetail with our 
Newtonian intuition. However, the successes of the Schwarzschild metric 
notwithstanding, it was pointed out by Eddington in the very early days of
general relativity that these same tests could also be met should some higher
derivative of $R^{\mu \nu}$ vanish rather than $R^{\mu \nu}$ itself. Eddington
thus raised the issue of how unique the Einstein equations actually are, and
despite the longevity and seriousness of the question, the relativity community 
seems to have somewhat ignored the issue, as well as the directly related one
we noted above, namely that covariance alone is not sufficient to select 
between metric theories of differing order. Recognizing the need to find
a principle which would enable us to make such an unambiguous selection,
Mannheim and Kazanas \cite{Mannheim1990}, \cite{MannheimandKazanas1989} 
suggested a candidate principle, namely conformal invariance (viz. invariance 
of the geometry under any and all local stretchings of the form 
$g_{\mu \nu}(x) \rightarrow \Omega^2(x) g_{\mu \nu}(x)$), the local scale 
invariance now thought to be enjoyed by the other fundamental interactions, the 
strong, the electromagnetic, and the weak, theories all of whose coupling 
constants are dimensionless and all of whose mass scales are 
dynamical.\footnote{Independent of one's views regarding conformal invariance 
itself, readers who object to it are still not free to use the Einstein 
equations until they come up with some other candidate principle which would 
then uniquely select out Eq. (\ref{(3)}), this being the challenge to the 
standard theory raised in \cite{Mannheim1994} and still not yet responded to.
Moreover, given the fact that the great appeal of Einstein gravity is that
it provides an explanation for the equality of the inertial and gravitational 
masses by means of a fundamental principle, it is thus quite extraordinary that
the choice of the particular equations of motion of Eq. (\ref{(3)}) would then  
be made without the analogous guidance of some other comparable underlying 
fundamental principle.} In conformal 
gravity (whose action is the conformal invariant 
$I_W=-\alpha \int d^4x (-g)^{1/2} C_{\lambda\mu\nu\kappa} 
C^{\lambda\mu\nu\kappa}$ where $C^{\lambda\mu\nu\kappa}$ is the conformal 
Weyl tensor and $\alpha$ is a purely dimensionless coupling constant) the 
equations of motion which are to replace the Einstein equations take the form
\begin{eqnarray}
 g^{\mu\nu}(R^{\alpha}_{\phantom{\alpha}\alpha})   
^{;\beta} _{\phantom{;\beta};\beta}/2                                             
+ R^{\mu\nu;\beta}_{\phantom{\mu\nu;\beta};\beta}                     
 -R^{\mu\beta;\nu}_{\phantom{\mu\beta;\nu};\beta}                        
-R^{\nu \beta;\mu}_{\phantom{\nu \beta;\mu};\beta}                          
 - 2R^{\mu\beta}R^{\nu}_{\phantom{\nu}\beta}                                    
+g^{\mu\nu}R_{\alpha\beta}R^{\alpha\beta}/2 
\nonumber \\
 -2g^{\mu\nu}(R^{\alpha}_{\phantom{\alpha}\alpha})          
^{;\beta}_{\phantom{;\beta};\beta}/3                                              
+2(R^{\alpha}_{\phantom{\alpha}\alpha})^{;\mu;\nu}/3                             
+2 R^{\alpha}_{\phantom{\alpha}\alpha} R^{\mu\nu}/3                               
-g^{\mu\nu}(R^{\alpha}_{\phantom{\alpha}\alpha})^2/6                   
= T^{\mu \nu}/4\alpha 
\label{(5)}
\end{eqnarray}
and admit of the Schwarzschild vacuum solution just as Eddington had warned us.
However, because they are fourth order rather than second order equations
these equations admit of other static, spherically symmetric vacuum solutions as
well, with the most general such one being found to take the form 
\cite{MannheimandKazanas1989} 
\begin{equation}
-g_{00}= 1/g_{rr}=1-\beta(2-3 \beta \gamma )/r - 3 \beta \gamma          
+ \gamma r - kr^2 
\label{(6)}
\end{equation}
where $\beta$, $\gamma$, and $k$ are integration constants.\footnote{This same 
solution was also found by Riegert \cite{Riegert1984}, and is contained in 
his PhD thesis (University of California at San Diego, 1986) in which conformal 
gravity is studied quite extensively, particularly with regard to its quantum 
aspects. This thesis also provides some quite extensive references to earlier 
conformal gravity work.} The metric of Eq. (\ref{(6)}) thus both contains and 
generalizes the Schwarzschild solution of standard gravity, recovering the 
standard Newton-Schwarzschild phenomenology when the $\gamma$ and $k$ terms are 
sufficiently small, while departing from it at sufficiently large enough 
distances through the presence of noneother than a linearly rising
potential,  a potential which is of just the type suggested by the 
phenomenological analysis we presented above 
in Sec. (2).  However, unlike the Schwarzschild solution, the metric of Eq. 
(\ref{(6)}) is not asymptotically flat, and in order to explore the 
implications of this fact for matter sources, it is necessary to relate the 
coefficients of this metric to interior properties of the source.
  
To this end we note that in a static, spherically symmetric geometry the 
equations of motion of Eq. (\ref{(5)}) 
remarkably reduce \cite{Mannheim1991}, \cite{MannheimandKazanas1994}
without any approximation whatsoever to           
\begin{equation}
\nabla^4 B(r) = (rB)^{\prime \prime \prime \prime}/r=f(r)
\label{(7)}
\end{equation}
where $B(r)=-g_{00}(r)$ and where $f(r) = {3(T^0_{{\phantom 0} 0} - 
T^r_{{\phantom r} r})/4\alpha B(r)}$
to thus yield a fourth order Poisson equation rather than the familiar second 
order one of the standard theory. In its turn Eq. (\ref{(7)}) is readily 
integrated and yields \cite{Mannheim1991}, \cite{MannheimandKazanas1994}
(see also \cite{PerlickandXu1995})  
\begin{equation}
B(r>R_0) =- {1 \over 6r} \int_{0}^{R_0} dr^{\prime} f(r^{\prime}) 
{r^{\prime}}^4
- {r \over 2} \int_{0}^{R_0} dr^{\prime} f(r^{\prime}) {r^{\prime}}^2
\label{(8)}
\end{equation}
as its exact exterior solution (up to an uninteresting $w-kr^2$ particular 
integral term). With the recovering of the Newtonian potential in Eq. 
(\ref{(8)}) we thus, in principle at least, divorce the Newtonian 
potential from the second order Poisson equation, to show that the second order 
Poisson equation is thus only sufficient to give the Newtonian potential but not 
in fact necessary (a state of affairs which will persist until some fundamental 
principle is identified which uniquely selects out the standard Newton-Einstein 
theory). Additionally from Eq. (\ref{(8)}) we see that our desired linearly rising 
potential term does indeed couple to matter, to show that in principle a local 
matter distribution of finite radius $R_0$ can in fact have a global effect 
at infinity, our Newtonian intuition notwithstanding.\footnote{In passing we note 
that since $-r/8\pi $ is both the Green's function and output potential to the 
fourth order Poisson equation, it plays the same role for $\nabla^4$ which 
$-1/4\pi r$ plays for the $\nabla^2$ operator.} In this way then higher order 
gravity theories thus become global. Having now motivated such rising potentials, 
we turn now to explore their possible observational implications.

In order to analyze dynamics based on rising potentials, beyond the issue of the 
fact that the form of the potential is different from that of the Newtonian one, 
the more fundamental difference is that in the same way that a given source now 
has dynamical implications at large distances, all the farthest sources from it 
then also have an effect on it, so it is now no longer possible to ignore the 
rest of the universe. Thus, for galactic rotation curves for instance, we need to 
examine the gravitational effect due to the linear potentials of the stars and 
gas within each galaxy, and also we need to examine the gravitational effect of 
the Hubble flow of the rest of the universe on motions within individual 
galaxies.\footnote{Since the rest of universe is only known to decouple in 
Einstein gravity when it is static and spherically symmetric, even for Einstein 
gravity an expanding, comoving universe could in principle affect motions within
individual galaxies, our Newtonian intuition notwithstanding.}

As regards first the linear potential contribution within a given galaxy, for 
weak gravity it is readily calculated in exactly the same manner as the net 
Newtonian potential contribution of a galaxy. Thus integrating the individual 
stellar potentials   
\begin{equation}
V^{*}(r)=-\beta^{*}c^2/r+
\gamma^{*}c^2r/2
\label{(9)}
\end{equation}
of Eqs. (\ref{(6)}) and (\ref{(8)}) (we redefine the coefficient of the stellar
Newtonian term as $\beta^{*}$) over an infinitesimally thin galactic optical disk 
with luminous surface matter distribution $\Sigma(R)=\Sigma_0$exp$(-R/R_0)$ and 
total number of stars $N^{*}=2 \pi \Sigma_0 R_0^2$ in the standard 
non-relativistic weak gravity way, then yields \cite{Mannheim1993a} the 
centripetal acceleration 
\begin{equation}
v^2/R=g_{gal}^{lum}=
g_{\beta}^{lum}+g_{\gamma}^{lum} 
\label{(10)}
\end{equation}
where
\begin{equation}
g_{\beta}^{lum}=
(N^{*}\beta^{*} c^2 r/2R_0^3)[I_0( r/2R_0)K_0( r/2R_0)-
I_1( r/2R_0)K_1(r/2R_0)]
\label{(11)}
\end{equation}
and
\begin{equation}
g_{\gamma}^{lum}=(N^{*}\gamma^{*} c^2r/2R_0)I_1( r/2R_0)K_1( r/2R_0)
\label{(12)}
\end{equation}
to give a net acceleration which behaves asymptotically as
$(v^2/c^2R)_{last}=\gamma^{*}N^{*}/2 +\beta^{*}N^{*}/R^2$. We thus nicely generate 
a $\gamma^{*}N^{*}/2$ term (an explicitly extensive function of the total 
amount of visible matter $N^{*}$ within each galaxy) just as needed for our 
phenomenological Eq. (\ref{(2)}). However, we  still lack
the $\gamma_0/2$ term which is also phenomenologically required. Since this
latter acceleration term is independent of the mass of a given galaxy, and since 
its numerically fitted value given in Sec. (2) is of order the inverse Hubble 
radius, we can thus anticipate that it must not arise from within a galaxy at all 
but must instead come from the global Hubble flow of the universe itself.  

In order to try to evaluate any possible contribution of the Hubble flow to
individual motions within galaxies, we are first faced with the difficulty that
in general (and in standard Einstein gravity in particular) trying to rewrite the 
comoving Robertson-Walker metric (a cosmological solution to both Einstein gravity
and to conformal gravity \cite{Mannheim1992}, \cite{Mannheim1995b}) in 
Schwarzschild coordinates is extremely complicated and anyway not particularly 
illuminating. However, this task is greatly simplified in conformal gravity 
because of its additional underlying conformal invariance. Specifically, it is 
found \cite{MannheimandKazanas1989} that the general coordinate transformation
\begin{equation} 
r=\rho/(1-\gamma_0 \rho/4)^2~~~,~~~ 
t = \int d\tau / R(\tau)
\label{(13)}
\end{equation} 
effects the metric transformation
\begin{eqnarray}
(1+\gamma_0 r)c^2dt^2-{dr^2 \over (1+\gamma_0 r)}-r^2d\Omega \rightarrow
\nonumber \\
{(1+\rho\gamma_0/4)^2 \over R^2(\tau)(1-\rho\gamma_0/4)^2}
\left(c^2 d\tau^2 - {R^2(\tau) (d\rho^2 + \rho^2 d\Omega)\over
(1-\rho^2\gamma_0^2/16)^2}
 \right) 
\label{(14)}
\end{eqnarray}
to yield a metric which we recognize as being a conformally transformed 
Robertson-Walker metric. Now while only the Robertson-Walker metric itself 
is a cosmological solution to Einstein gravity, in conformal gravity not only is 
Robertson-Walker an allowed cosmological solution, but so is any metric which is 
conformal to it. Thus we see that a static, Schwarzschild coordinate linear 
potential metric is coordinate equivalent to the conformally equivalent class of 
allowed conformal gravity cosmologies with scale factor $R(\tau)$ and (explicitly 
negative) 3-space scalar curvature $k=-\gamma_0^2/4$.\footnote{In passing we note 
that in the cosmology discussed in \cite{Mannheim1992}, \cite{Mannheim1995b} an 
open universe with explicitly negative $k$ was in fact realized, with such a 
universe being found to not suffer from the flatness problem found in the 
standard cosmology.} Now, as regards this equivalence, we 
note that in a geometry which is both homogeneous and isotropic about all 
points, any observer can serve as the origin for the coordinate $\rho$; thus 
in his own local rest frame each observer is able to make the above general 
coordinate transformation with the use of his own particular $\rho$. 
Moreover, since the observer is also free in conformal gravity to make 
arbitrary conformal transformations as well, that observer will then be able  
to see the entire Hubble flow appear in his own local static coordinate 
system as a universal linear potential with a universal acceleration 
$\gamma_0c^2/2$ coming directly from the spatial curvature of the universe 
(a quantity which incidentally is nicely time independent unlike the time 
dependent Hubble parameter itself). Since the internal orbital motions of the 
stars and gas can be discussed in each galaxy's own rest frame, we thus find 
that in each such rest frame, each orbiting particle in that specific galaxy 
will then precisely see the overall Hubble flow acting as a local static 
universal linear $\gamma_0 r$ potential just as desired. (Of course, in the rest 
frame coordinate system associated with a given galaxy, any observer on any other 
galaxy will see an extremely complicated non-static geometry. However, in his own 
rest frame that observer is still free to use Eq. (\ref{(14)}) with the selfsame 
universal acceleration to describe the internal motions in his own galaxy, which 
is just what is needed for that galaxy's rotation curve.) We thus establish a 
cosmological origin for the universal $\gamma_0c^2/2$ acceleration needed for 
Eq. (\ref{(2)}), while also identifying a crucial difference between 
relativistic and non-relativistic reasoning. Specifically, in strictly 
Newtonian physics the only effect of any background would be to put tidal 
forces on individual galaxies, forces that would not account for the 
rotational motions of stars and gas but only to a departure therefrom. 
Relativistically however, since the background produces an effect at 
the center of each galaxy, the background therefore contributes to the 
explicit rotational motions of the stars themselves, to thus yield a 
previously unappreciated but nonetheless apparently quite general 
consequence of curvature which enables the stars and gas in galaxies to serve as
test particles for probing the Hubble flow.

In order to now combine the above local and global linear potentials we need to
embed each local galaxy into the global Hubble flow and solve the gravitational 
equations of motion of Eq. (\ref{(5)}) in the presence of 
$T^{\mu \nu}_{local}+T^{\mu \nu}_{global}$.  Given the fact that gravity is weak 
within galaxies, we shall as a first approximation 
simply add the local and global metrics given above in Eqs. (\ref{(11)}),
(\ref{(12)}) and (\ref{(14)}) (it is the very presence of 
$T^{\mu \nu}_{local}$ and its associated local geometry (viz. standard 
static Schwarzschild coordinates) which dictates the appropriate general 
coordinate transformation needed for Eqs. (\ref{(13)}) and  
(\ref{(14)})), to yield the total weak gravity acceleration 
\begin{equation}
v^2/R=g_{tot}=g_{\beta}^{lum}+g_{\gamma}^{lum}+\gamma_0 c^2/2
\label{(15)}
\end{equation}
which can now be directly applied to data. With $\gamma_0$ and $\gamma^{*}$ 
taking the fixed numerical values given earlier in Sec. (2), the fits reduce to 
just one free parameter per galaxy, viz. the standard optical disk mass to 
light ratio (or equivalently the total amount of stars and gas per galaxy, 
$N^{*}$, in solar mass units). Since, unlike dark matter theory, our theory 
is based on parameters with an absolute scale, it is thus very sensitive to 
distance determinations to galaxies. Consequently, we first calculate the 
total velocity predictions (the dotted curves) in Fig. (1) using the 
distances quoted in \cite{BegemanBroeilsandSanders1991}. 
Then, again following  \cite{BegemanBroeilsandSanders1991} we allow 
for typical uncertainties in the adopted distances to give modest distance 
shifts of up to $\pm 15\%$ or so. (While larger shifts can actually 
improve the fits a little in some cases, we have not allowed for shifts of 
more than this except for NGC 1560 for which a distance estimate of 3.7 Mpc 
($+23\%$) has actually been reported in the literature.) With the indicated 
percentage shifts in adopted distance, with the fitted $M/L$ ratios listed in 
Table (1), and with $g_{gal}^{lum}$ of Eq. (\ref{(10)}) being calculated solely 
from the known luminous galactic matter (viz. stars and gas), we then obtain (the 
details of the fitting are given in \cite{Mannheim1996a}) the full 
curve fits of Fig. (1), with the dashed and dash-dotted curves showing the 
velocities that the Newtonian $g_{\beta}^{lum}$ and linear $g_{\gamma}^{lum}
+\gamma_0c^2/2$ terms would separately produce. No dark matter is assumed, and 
as we can see from the fits, none would appear to be needed. Despite the fact 
that our model is a highly constrained one with very few free parameters, it 
nonetheless appears to have captured the essence of the data (our fits have 
smoothed out some of the structure in the data since we treat the radial
dependence of the optical disks as single exponentials for simplicity), and 
phenomenologically our fitting 
would thus appear to be competitive with that of both the standard dark matter 
model and the MOND \cite{Milgrom1983} alternative. Moreover, if our theory is in 
fact correct, then it provides us with an actual determination of the scalar 
curvature of the universe, something which years of intensive work 
has yet to accomplish in the standard theory.  
     
As we can see from Fig. (1), at the shifted adopted distances $(v^2/c^2R)_{last}$ 
is indeed remarkably well fitted by 
$\gamma_0/2+\gamma^{*}N^{*}/2+\beta^{*}N^{*}/R^2$ (this being the asymptotic limit 
of $g_{tot}/c^2$), to thus recover Eq. (\ref{(2)}) for each and every galaxy in 
our sample; and we see that even while the quantity $\gamma^{*}N^{*}/2$ does vary
enormously with luminosity over our sample, nonetheless the $\gamma_0/2$ term 
overwhelms it in all but the largest galaxies, so that $(v^2/c^2R)_{last}$ 
only shows a mild (but nonetheless significant) dependence on galactic 
mass.\footnote{With the ratio 
$(\gamma_0/2+\gamma^{*}N^{*}/2)/(\beta^{*}N^{*}/R^2)$ decreasing as $N^{*}$ 
increases, we thus see that departures from the luminous Newtonian expectation are 
the biggest in the low luminosity galaxies. This then naturally parallels the 
analogous trend found phenomenologically for the luminosity dependence of the 
dark to luminous mass ratio in dark matter models to which we referred earlier. 
In passing we also recall that even though it is not all that clear as to just
how dark matter is in fact clever enough to always manage to appropriately match 
itself to the luminous matter content of individual galaxies, the fact that 
there is such a trend with luminosity has nonetheless spawned a folklore theorem 
that it is impossible to replace dark matter by any new theory with only one new 
scale. As Eq. (\ref{(2)}) now shows, it is, however, possible to replace dark 
matter by a theory with two.} Since we find that the same $N^{*}$ as extracted out 
from $(v^2/c^2R)_{last}$ also serves to accurately normalize the luminous 
Newtonian contribution in the inner rotation region where it is dominant, we thus 
see that the parameters extracted out from $(v^2/c^2R)_{last}$ alone then provide 
for a complete accounting of the entire rotation curves with no further adjustment
of parameters being needed at all, a fitting which stands in sharp contrast to 
dark matter halo fitting where no less than 22 halo parameters are used to fit 
the data set of Fig. (1).         

While we have thus made a first case for a possible role for cosmology in
elucidating the systematics of galactic rotation curves, (the data of Fig. (1) 
are certainly not rejecting the idea), nonetheless this idea has to be explored 
further especially on distance scales larger than galactic where background 
cosmological effects should be even more prominent.\footnote{A first application 
\cite{KnoxandKosowsky1993}, \cite{ElizondoandYepes1994} of conformal gravity to 
cosmological nucleosynthesis has actually encountered a possible problem with
conformal cosmology so far having difficulty generating sufficient primordial 
deuterium.} In conclusion then, we would like to state that the conformal gravity 
theory would appear capable of explaining the general systematics of galactic 
rotation curves in a completely natural manner, and that our study suggests that 
rising rather than flat rotation curves is actually the paradigm, with the 
luminous Newtonian contribution having inadvertently masked that fact in the 
higher luminosity galaxies. Moreover, through the cosmological connection we 
have presented, we believe we have made a case for the existence of a 
universal linear potential associated with the cosmological Hubble flow (and even 
more generally a case against the widespread belief that whatever is responsible
for the rotation curve discrepancies is itself a purely local galactic 
phenomenon), an intriguing possibility which appears to enable us to circumvent 
the need for galactic dark matter.  

\section{Global aspects of wave propagation in a gravitational field}

Having now explored some global gravitational aspects on large astrophysical 
distance scales, we turn next to an initially unlikely place to identify another
one, namely a neutron interferometry experiment performed in a terrestrial 
laboratory. In a landmark series of experiments \cite{Overhauser1974}, 
\cite{Colella1975} Colella, Overhauser and Werner (COW) and their subsequent 
collaborators (see e.g. Refs. \cite{Greenberger1979}, \cite{Werner1994} for 
overviews) detected the modification of the phase of a neutron beam as it 
traverses the earth's gravitational field, to thus realize the first 
experiment which involved both quantum mechanics and gravity. A typical generic 
experimental set up is shown in the schematic Fig. (2) in which a neutron 
beam from a reactor is split by Bragg or Laue scattering at point $A$ into a 
horizontal beam $AB$ and a vertical beam $AC$ (we take the Bragg angle to be 
$45^{\circ}$ for simplicity and illustrative convenience in the following), with 
the subsequent scatterings at $B$ and $C$ then producing beams which Bragg scatter 
again at $D$, after which they are then detected. If the neutrons arrive at $A$ 
with velocity $v_0$ (typically of order $10^5$ cm sec$^{-1}$) and $ABCD$ is a 
square of side $H$ (typically of order a few centimeters), then the phase 
difference between the two split beams is given by
\begin{equation}
\phi_{COW}=\phi_{ACD}-\phi_{ABD}=-mgH^2/\hbar v_0 
\label{(16)}
\end{equation}
to lowest order in the acceleration $g$ due to gravity 
\cite{Overhauser1974}, and is actually observable despite the weakness of gravity 
($mgH/(mv_0^2/2)\simeq 10^{-7}$ in typical experimental conditions), since even 
though $\int \bar{p} \cdot d\bar{r}$ only differs by the minute amount 
$m(v_{CD}-v_{AB})H=-mgH^2/v_0 \simeq 10^{-26}$ erg sec between the $CD$ and $AB$ 
paths, nonetheless this quantity is not small compared to Planck's constant, to 
thus give an detectable fringe shift (compared to traversing the same 
interferometer with the plane of $ABCD$ horizontal) even for $H$ as small as a 
few centimeters. Quantum mechanical interference based experiments thus provide 
a level of sensitivity for probing background classical gravity way in excess of 
that achievable in classical Newtonian gravity based laboratory experiments of 
the same characteristic dimensions.
    
The detected COW phase is extremely intriguing for two reasons. First, it shows 
that it is possible to distinguish between different paths which have common end 
points, with the explicit global ordering in which the horizontal and vertical 
sections are traversed leading to observable consequences. And second, it yields 
an answer which explicitly depends on the mass of the neutron even while the 
classical neutron trajectories (viz. the ones explicitly followed by the centers 
of the wave packets of the quantum mechanical neutron beam) of course do not. 
The COW result thus invites consideration of whether the detected ordering is 
possibly another example of the kind of global topological effects which are 
characteristic of quantum mechanics, and of whether quantum mechanics actually 
respects the equivalence principle. As we shall see, the ordering effect is in 
fact already present in the propagation of classical light waves around the 
interferometer with classical light undergoing interference in a classical 
gravitational background, to show that classical gravity is intrinsically global.
Once classical waves undergo interference in some given background, it then
follows from wave particle duality that quantum mechanical matter waves must 
inherit the same interference pattern in the same background, so that the neutron 
COW phase derives from global classical gravity rather than from topological 
quantum mechanics. Also we will see that this same wave particle duality will 
enable us to establish that the mass dependence of the neutron beam COW phase is 
purely kinematic with the equivalence principle then continuing to hold in the 
presence of quantum mechanics.

To address these issues specifically we have found it convenient to carefully
track a light wave as it traverses the interferometer (the discussion parallels
that of \cite{Mannheim1996b} which monitors the motion of the neutron 
around the loop). Since the polarization of the light wave is not relevant to our
considerations, it suffices to look at solutions to the scalar Klein-Gordon 
equation $\phi^\mu _{\phantom{\mu};\mu}=0$ ($\phi^{\mu}$ denotes 
$\partial\phi/\partial x_{\mu}$) in generic background fields of the form 
$d\tau^2=B(r)c^2dt^2-dr^2/B(r)-r^2d\Omega$ where $B(r)=1-2MG/c^2r$ near the 
surface of the earth (in both Einstein gravity and conformal gravity since the
linear potential contribution is completely negligible here). On defining  
$\phi(x)=$exp$(iT(x))$ we find that the phase of the
wave obeys $T^{\mu}T_{\mu}=i T^{\mu}_{\phantom{\mu};\mu}$. In the eikonal or 
ray approximation the $i T^{\mu}_{\phantom{\mu};\mu}$ term can be dropped, 
so that the phase $T(x)$ is then seen to obey the classical massless particle 
Hamilton-Jacobi equation $T^{\mu}T_{\mu}=0$, an equation whose solution is the 
stationary classical action between relevant end points. In the eikonal  
approximation we can also identify $T^{\mu}$ as the wavefront normal 
$dx^{\mu}/dq$ where $q$ is a convenient affine parameter which can be 
used to measure distances along ray trajectories. On introducing the wave number 
$k^{\mu}=dx^{\mu}/dq$, we can then set $T(x)=\int k_{\mu}dx^{\mu}$, to thus 
recover the standard expression for the phase of a classical wave. Additionally, 
since the covariant differentiation of the Hamilton-Jacobi equation yields 
$(T^{\mu}T_{\mu})_{;\nu}=2T^{\mu}T_{\mu;\nu}                                
=2T^{\mu}(T_{\nu;\mu}+\partial_{\mu}T_{\nu}-\partial_{\nu}T_{\mu})            
=2T^{\mu}T_{\nu;\mu}=2k^{\mu}k_{\nu;\mu}=0$, the
identification $k^{\mu}=dx^{\mu}/dq$ then yields the massless particle 
geodesic equation \cite{Mannheim1993b}. In the eikonal approximation then we see 
that light rays travel on the curved space geodesics. However, while the 
identification $T^{\mu}=k^{\mu}$ nicely puts $k^{\mu}$ on the light cone, it 
also causes 
$\int k_{\mu}dx^{\mu}=\int (dx_{\mu}/dq)dx^{\mu}$ to vanish identically and thus 
not change along trajectories, to initially suggest that there is no detectable 
interference. However, since the same argument would lead us to the conclusion
that a Young double slit experiment with classical light would also not yield 
any interference pattern, it is thus instructive to explicitly identify why it is 
that the double slit experiment does in fact display interference.

In a double slit experiment with classical light, light from a source $S$ goes 
through slits $Q$ and $R$ (see Fig. (3)) to form an interference pattern
at points such as $P$, with the distance $\Delta x=QT$ representing the 
difference in path length between the two beams. Given this path difference, 
the phase difference between the two beams is usually identified as $k\Delta x$, 
from which an interference pattern is then readily calculated. However, because 
of this path difference, the $SQP$ ray takes the extra time $\Delta t 
=\Delta x/c$ to get to $P$, to thus give a net change in the phase of the $SQP$ 
beam equal to  $k\Delta x-\omega \Delta t$, a quantity which actually vanishes for 
rays on the light cone, just as had been noted above. The relative phase of the 
two light rays in the double slit experiment thus does not change at all as the 
two beams traverse the interferometer, and cannot thus be the cause of the 
interference pattern. However, because of this time delay, the $SRP$ beam actually 
arrives at $P$ at the same time as an $SQP$ beam which had left the source a time 
$\Delta t$ earlier. Thus if the source is coherent over these time scales, then
relative to the $SRP$ beam the $SQP$ beam then carries an additional 
$+\omega \Delta t$ phase from the very outset, and it is this specific additional 
phase which is then responsible for the detected interference.\footnote{Since 
this phase is equal and opposite to the $-\omega \Delta t$ phase which the same 
beam acquires during its propagation to $P$, the phase $k\Delta x$ can still serve 
as the final observable phase difference for purposes of actually calculating the 
explicit interference pattern, though this specific phase is non-zero only if 
there is in fact a time delay.} We thus see that the double slit device itself 
actually produces no phase change for light. Rather, the choice of point $P$ on 
the screen is a choice which selects which time delays at the source are relevant 
at each $P$, with any general interference pattern for light thus not only always 
involving the time delay at the source, but also in fact always requiring one.
Consequently, we now need to determine whether light waves also experience any 
analogous time delay when they traverse an interferometer placed in a 
gravitational field.
  
In order to determine whether there is any such time delay, we need to determine
the light ray geodesics in the gravitational field of the earth. To this end it is 
convenient to rewrite the Schwarzschild metric in terms of a Cartesian coordinate 
system $x=r sin \theta cos \phi,~y=r sin \theta sin \phi,~z=r cos \theta -R$ 
erected at a point on the surface of the earth. With $z$ being normal to the 
earth's surface, to lowest order in $x/R,~y/R,~z/R,~MG/c^2R$ (where $M$ is the 
mass of the earth and $R$ its radius) the Schwarzschild metric is then found 
\cite{Moreau1994} to take the form
\begin{equation}
d\tau^2=f(z)c^2dt^2-dx^2-dy^2-dz^2/f(z)
\label{(17)}
\end{equation}
where $f(z)=1-2MG/c^2R+2gz/c^2$ and where $g$ denotes $MG/R^2$. Since the 
non-relativistic geodesics associated with the metric of Eq. (\ref{(17)}) are 
given by $\ddot{x}=0,~\ddot{y}=0,~\ddot{z}=-g$, we see that this 
metric nicely describes a constant gravitational acceleration. Moreover,
since the coordinate transformation \cite{Greenberger1979} 
\begin{equation}
ct^{\prime} 
=c^2sinh(gt/c) f^{1/2}(z)/g~~~,~~~z^{\prime}=c^2(cosh(gt/c)f^{1/2}(z)
-f^{1/2}(0))/g 
\label{(18)}
\end{equation}
(a transformation which reduces to
\begin{equation}
ct^{\prime} 
=ct(1-gR/c^2)+tgz/c~~~,~~~z^{\prime}=z(1+gR/c^2)+gt^2/2-gz^2/2c^2
\label{(19)}
\end{equation}
to lowest non-trivial order in $g$) transforms the metric of Eq. (\ref{(17)}) 
into the flat Cartesian metric 
$d\tau^2=c^2dt^{\prime 2} -dx^2-dy^2-dz^{\prime 2}$, we see that the metric of 
Eq. (\ref{(17)}) thus nicely incorporates the equivalence principle, not only
for material particles, but necessarily also for light (explicitly because 
of covariance in fact). For the metric of Eq. (\ref{(17)}) the Hamilton-Jacobi
equation takes the form of the light cone condition
\begin{equation}
f(z)(k^0)^2-(k^1)^2-(k^2)^2-(k^3)^2/f(z)=0
\label{(20)}
\end{equation}
while the massless particle geodesic equations take the form 
\begin{eqnarray}
k^0=cdt/dq=\alpha_0/f(z)~~~,~~~k^1=dx/dq=\alpha_1
\nonumber \\
k^2=dy/dq=\alpha_2~~~,
~~~k^3=dz/dq=(\alpha_0^2-(\alpha_1^2+\alpha_2^2)f(z))^{1/2}
\label{(21)}
\end{eqnarray}
where the $\alpha_i$ are integration constants. From these equations we 
recognize the existence of the gravitational frequency shift (since $k^0$ 
depends on $z$), the gravitational time delay ($dz/dt$ depends on $z$), and the 
gravitational bending of light ($dz/dx$ depends on $z$ if $\alpha_1\neq 0$), with
all of these effects thus participating in the motion of a classical light wave 
around the $ABCD$ interferometer loop. 

While the geodesic equations of Eq. (\ref{(21)}) enable us determine the 
trajectories of the light rays in between the various scattering surfaces, in 
order to be able to do the complete calculation we also need to know the rules for 
the scattering of light at the individual crystal surfaces, i.e. we need to 
determine the Bragg scattering rules in curved space, something which we can do
via a sequence of coordinate transformations. Specifically, consider a ray which
takes a time $T$ to travel vertically from $A$ to $C$ in the rest frame
of the interferometer in the presence of the curved space metric of Eq. 
(\ref{(17)}). According to Eq. (\ref{(21)}) it arrives at $C$ with four momentum
$k^{\mu}_{in}=(\alpha_0/f(H),0,0,\alpha_0)$. From the point of view of the 
accelerating flat space observer associated with Eq. (\ref{(18)}), to lowest order 
in $g$ the point $C$ is moving upward with velocity $v=gt^{\prime}=gT=gH/c$ at 
that instant, and the wave has momentum 
$k^{\prime \mu}_{in}=\alpha_0(1+gR/c^2)(1,0,0,1)$ 
to this same order. To an observer who is moving vertically downward with this same 
velocity $v$ at that same instant, the scattering crystal appears to be at rest 
and the wave appears to have momentum 
$k^{\prime \prime \mu}_{in}=\alpha_0(1+gR/c^2)(1-v/c)(1,0,0,1)$ to lowest order. In this 
frame the wave undergoes standard rest frame flat space Bragg scattering and is 
thus reflected into the horizontal direction with momentum 
$k^{\prime \prime \mu}_{out}=\alpha_0(1+gR/c^2)(1-v/c)(1,1,0,0)$. Applying next 
a Lorentz transformation with velocity $v$ vertically upwards transforms back to 
the coordinate system of Eq. (\ref{(18)}) where the outgoing wave 
then has momentum 
$k^{\prime \mu}_{out}=\alpha_0(1+gR/c^2)(1-v/c,1-v/c,0,v/c)$ to lowest 
order in $g$. Finally, returning to the original coordinate system of 
Eq. (\ref{(17)}) yields a wave with momentum 
$k^{\mu}_{out}=(\alpha_0/f(H),\alpha_0/f^{1/2}(H),0,0)$ which is nicely seen to 
obey the light cone constraint of Eq. (\ref{(20)}) to this order. In the 
original coordinate system we thus see that the general Bragg scattering rule in
curved space is that the wave undergoes no change in the magnitude of $k^0$ in a 
Bragg scattering, that it emerges with an angle of reflection 
equal to the angle of incidence, and that the magnitudes of the spatial 
components of the outgoing momentum take whatever values are needed to keep the 
outgoing wave on the light cone. Armed with this result we can now finally
track light rays around the interferometer loop.

Explicit calculation (to lowest non-trivial order in $g$ throughout) indicates 
that the light ray which goes up vertically from $A$ to $C$ sets out with 
momentum $(\alpha_0/f(0),0,0,\alpha_0)$ and arrives at $C$ with a 
momentum $(\alpha_0/f(H),0,0,\alpha_0)$ after a travel time $t(AC)$ given by
\begin{equation}
t(AC)=H(1-gH/c^2+2gR/c^2)/c
\label{(22)}
\end{equation}
\noindent On scattering at $C$ the wave is then reflected so that it starts 
off toward $D$ with momentum $(\alpha_0/f(H),\alpha_0/f^{1/2}(H),0,0)$. On its 
flight gravitational bending causes it to dip slightly so that it arrives not at
$D$ but rather at the point $D_1$ with coordinates $(H-gH^2/2c^2,H-gH^2/2c^2)$, 
with the $CD_1$ segment taking a time $t(CD_1)$ 
given by
\begin{equation}
t(CD_1)=H(1-3gH/2c^2+gR/c^2)/c
\label{(23)}
\end{equation}
\noindent
The wave which starts horizontally from $A$ with momentum 
$(\alpha_0/f(0),\alpha_0/f^{1/2}(0),0,0)$ arrives not at $B$ but at the 
point $B_1$ with coordinates $(H-gH^2/2c^2,-gH^2/2c^2)$ 
and with a momentum 
$(\alpha_0/f(0),\alpha_0/f^{1/2}(0),0,
-\alpha_0 gH/c^2f^{1/2}(0))$. 
The $AB_1$ segment takes a time
$t(AB_1)$ given by
\begin{equation}
t(AB_1)=H(1-gH/2c^2+gR/c^2)/c
\label{(24)}
\end{equation}
\noindent
At $B_1$ the scattered wave sets off toward $D$ with momentum
$(\alpha_0/f(0), -\alpha_0 gH/c^2f(0),0,\alpha_0)$
and arrives not at $D$ or $D_1$ but rather at the point
$D_2$ with coordinates $(H-3gH^2/2c^2,H-3gH^2/2c^2)$, with the 
$B_1D_2$ segment taking a time
$t(B_1D_2)$ given by
\begin{equation}
t(B_1D_2)=H(1-2gH/c^2+2gR/c^2)/c
\label{(25)}
\end{equation}
\noindent
As regards the light wave path around the loop, we thus see that the small 
vertical $gH^2/2c^2$ dip during each of the two horizontal legs causes each ray 
to travel a distance $gH^2/2c^2$ less in the horizontal than it would have done 
in the absence of gravity. As regards the two vertical legs, we note that even 
though the $AC$ leg is completely in the vertical, since the $B_1D_2$ ray starts 
its leg with a small horizontal momentum (which it acquired because of 
the small dip in the prior $AB_1$ leg), during this leg the ray changes 
its horizontal coordinate by an amount $gH^2/c^2$, thereby causing it to reach 
$D_2$ after having also traveled a distance $gH^2/c^2$ less in the vertical 
than it would travel in the $AC$ leg. Consequently, there is a spatial 
offset $(gH^2/c^2,~gH^2/c^2)$ between $D_1$ and $D_2$ which means that these
particular beams are not able to interfere. Moreover, we also
see that there is not anyway any time delay between the arrival of the two 
beams since $t(AC)+t(CD_1)-t(AB_1)-t(B_1D_2)= 0$.

Before discussing the issue of this spatial offset, it is instructive to ask
where the beams would have met had there been no third crystal at $D$ to get 
in the way. Explicit calculation shows that they would in fact have met at the 
asymmetric point $D_3$ with coordinates 
$(H-3gH^2/2c^2,H-gH^2/2c^2)$, with the $CD_3$ and $B_1D_3$
segments respectively taking the times $t(CD_3)$ and $t(B_1D_3)$ given by
\begin{equation}
t(CD_3)=H(1-5gH/2c^2+gR/c^2)/c~~~,~~~t(B_1D_3)=H(1-gH/c^2+2gR/c^2)/c
\label{(26)}
\end{equation}
\noindent 
The rays would thus have met at $D_3$ with time delay
$t(AC)+t(CD_3)-t(AB_1)-t(B_1D_3)= -2gH^2/c^3$, with both the time delay and the 
fact that the rays would meet at an asymmetric point rather than on the $AD$ axis 
thus directly revealing the explicit nature of the global ordering effect which	a 
background gravitational field has on a closed loop path.

As regards the spatial offset between the two beams, because of it the $AB_1D_2$ 
path interferes not with the $ACD_1$ path, but rather with the indicated nearby 
offset $A_1C_1D_2$ path, a very close by path which in fact is found to lie a 
distance $gH^2/c^2$ vertically below $AB$, an offset distance which is well 
within the resolution of any incident beam.  However, because of the spatial 
offset between $D_1$ and $D_2$, the $AB_1D_2$ path beam has to travel an extra 
horizontal distance $A_2A=gH^2/c^2$ to first get to the interferometer 
(to therefore provide an analog to the distance $\Delta x=QT$ in the double 
slit experiment, with $A_1$ and $A_2$ acting just like the pair of slits $Q$ 
and $R$). This spatial offset itself leads to a time delay $A_2A/c$, so that 
finally observable interference is produced,\footnote{Since the sign of the time 
is unchangeable by coordinate transformations, the fact of a time delay between 
the two beams at the source is thus a covariant indicator for interference.} 
with the two beams acquiring a net phase shift $-\alpha_0gH^2/c^2$ due entirely
to the $A_2A$ segment alone. On recognizing that $\alpha_0$ is 
the value of $k^0$ at $z=0$ we may set it equal to $2\pi/\lambda$ where 
$\lambda$ is the wavelength of the incident beam, to finally obtain for the 
phase shift $\Delta \phi_{CL}=-2\pi gH^2/\lambda c^2$ where $CL$ denotes 
classical light. Now while $H$ would have to be of the order of $10^5$ cm for 
$\Delta \phi_{CL}$ to actually be detectable in a Bragg scattering 
interferometer of the same sensitivity as the neutron COW 
experiment,\footnote{While this is still sizable for an interferometer, its 
interest lies in the fact that it allows us to detect, in principle at least, 
the gravitational bending of light using laboratory sized distance 
scales rather than solar system sized ones. Thus it would
also be of interest to see what dimension interferometer might serve as a 
gravitational wave detector.} nonetheless we can still identify this phase 
shift as an in principle, completely classical effect which reveals the 
intrinsically global nature of classical gravity.

Now that we have obtained $\Delta \phi_{CL}=-2\pi gH^2/\lambda c^2$ it is 
instructive to compare it with $\Delta \phi_{COW}$. If we introduce the neutron 
de Broglie wavelength $\lambda_n = h/mv_0$, we may rewrite the neutron 
$\Delta \phi_{COW}$ phase of 
Eq. (\ref{(16)}) in the form $-2\pi gH^2/\lambda_n v_0^2$. Comparison with 
$\Delta \phi_{CL}$ thus reveals a beautiful example of wave particle duality, 
with the quantum mechanical neutron matter wave inheriting its interference 
aspects from the behavior of the underlying classical wave. Thus even while 
$\Delta \phi_{COW}$ does depend on the mass of the neutron,\footnote{The reason 
why $\Delta \phi_{COW}$ actually depends on $m$ at all was given in 
\cite{Mannheim1996b} where it was shown that an analogous spatial offset 
$gH^2/v_0^2$ and time delay $gH^2/v_0^3$ occur for neutron beams, with the 
neutron COW phase then being given as the associated net change in the neutron 
action due to traversing the loop of Fig. (2). Specifically, it was noted that 
even while the stationary trajectories of the classical 
$-mc\int d\tau$ action (viz. the trajectories explicitly followed 
by the centers of the quantum mechanical neutron beam wave packets) are 
themselves totally independent of $m$ (the equivalence principle) to thus
give a spatial offset whose magnitude is independent of $m$, nonetheless 
the actual value of the classical action itself in any such trajectory does 
nonetheless depend on $m$, though only as a purely kinematic overall multiplying 
factor whose presence does not affect the position of the minimum of the action. 
And, even though the actual value of the stationary classical neutron point 
particle action is not observable classically, nonetheless it does become 
observable quantum mechanically as the phase of the neutron wave function, a 
phase whose overall normalization explicitly (kinematically) depends on $m$.} 
its dependence is strictly a passive, kinematic one with gravity only coupling 
to neutron matter waves via their de Broglie wavelengths. Thus the neutron mass  
has no dynamical consequences in Eq. (\ref{(16)}) with the explicit spatial 
offset and time delay needed to produce interference actually being completely 
independent of the mass of the neutron. Consequently, the COW experiment would 
appear to be completely compatible with the equivalence principle.

Now while we have shown that the COW phase does not derive from topological
quantum mechanical considerations, it is nonetheless somewhat puzzling that the	
resulting COW phase is proportional to the area enclosed by the $ABCD$ loop, 
since proportionality to the area is a characteristic topological signature.
However, in such topological cases, the area in question is typically  
related to the flux threaded by some electromagnetic field, whereas in the
gravitational case the gravitational field is parallel to the loop not
perpendicular to it, with there instead being a gravitational field gradient in 
the plane of the loop. In order to underscore this remark, it is instructive to 
consider the familiar Michelson Morley experiment, only with one of the 
interferometer arms 
vertical rather than horizontal. In such a situation the crystal at $B$ in 
Fig. (2) would be replaced by a vertical mirror and that at $C$ by a horizontal 
one, with the beams both splitting and recombining at $A$ to then be detected at 
a location below $A$. Such a device would then have no enclosed area. However, 
due to gravitational bending in the $AB$ and $BA$ legs, there would still be a 
spatial offset at $A$ between the returning $BA$ and $CA$ beams of order 
$gL^2/c^2$ where $L$ is the $AB$ length. Consequently, compared to a horizontal 
Michelson Morley interferometer, one with one arm vertical would undergo a 
detectable phase shift\footnote{To be precise, the gravitational bendings during 
the $AC$ and $CA$ legs which would occur when $AC$ is in the horizontal no 
longer occur when $AC$ is in the vertical, so they can then no longer offset the 
bendings which occur in the horizontal $AB$ and $BA$ 
legs.} of order $2\pi gL^2/\lambda c^2$, even though $L^2$ is not a measure of 
any relevant interferometer area. Rather, the bending distance in any arm is 
given by the length of that arm multiplied by the only available dimensionless 
perturbation parameter appropriate to the problem, viz $gL/c^2$, with the 
ensuing spatial offset $gL^2/c^2$ itself thus being quadratic in $L$ even while 
it has the dimension of length rather than of area. (Noting that since the $AB$ 
path sweeps out a small (triangular) area in going to $B$ and back, and that 
similarly the enclosed COW experiment loop differs slightly from the ideal 
$ABCD$ loop, to the extent that any specific area is relevant, it would be the 
one which represents the difference between the interferometer paths associated 
with the vertical and horizontal configurations.) Like the Michelson Morley 
case, the COW phase 
shift is thus explicitly seen to not be of topological origin, but rather, just 
like the impact of the Hubble flow on galactic rotation curves, is instead seen 
to be a further manifestation of the global nature of gravity, a thus 
characteristic feature of relativistic gravity which seems not to have been 
given all that much emphasis in the literature. This work has been supported in 
part by the Department of Energy under grant No. DE-FG02-92ER40716.00.

\vfill\eject

 \vfill\eject
\medskip
\noindent
{\bf Table (1)}
\medskip
$$
\begin{array}{cccccccc}

 Galaxy &  Distance & Luminosity & (v^2/c^2R)_{last} & Shift  &
(M/L) & \\

 {}& (Mpc) & (10^9L_{B \odot}) & (10^{-30}cm^{-1})  &
(\%)& 
(M_{\odot}L_{B \odot}^{-1})&  \\

{}&{}&{}&{}&{}&{} \\ 

 DDO\phantom{1}~154  &\phantom{0}3.80& \phantom{0}0.05 & 1.51& 
-11 & 0.71   \\

 DDO\phantom{1}~170  &          12.01& \phantom{0}0.16 & 1.63&
-07 & 5.36   \\

 NGC~1560&            \phantom{0}3.00& \phantom{0}0.35 & 2.70&
+23 & 2.01 &   \\

 NGC~3109&            \phantom{0}1.70& \phantom{0}0.81 & 1.98&
    & 0.01  &  \\

 UGC~2259&            \phantom{0}9.80& \phantom{0}1.02 & 3.85&
+15 & 3.62  &   \\

 NGC~6503&            \phantom{0}5.94& \phantom{0}4.80 & 2.14&
    & 3.00  &   \\

 NGC~2403&            \phantom{0}3.25& \phantom{0}7.90 & 3.31&
+15 & 1.76  &   \\

 NGC~3198&            \phantom{0}9.36& \phantom{0}9.00 & 2.67& 
-15 & 4.78  &   \\

 NGC~2903&            \phantom{0}6.40&           15.30 & 4.86&
+14 & 3.15  &   \\

 NGC~7331&                      14.90&           54.00 & 5.51& 
-16 & 3.03  &  \\

 NGC~2841&            \phantom{0}9.50&           20.50 & 7.25&   
    & 8.26  & 
 
\end{array}
$$

\noindent
{\bf Figure Captions}

\medskip
\noindent
Figure (1). The predicted rotational velocity curves associated with 
conformal gravity for each of the 11 galaxies in the sample. In each graph 
the bars show the data points with their quoted errors, the full curve 
shows the overall (adopted distance adjusted) theoretical velocity 
prediction (in km sec$^{-1}$) as a function of distance from the center of 
each galaxy (in units of $R/R_0$ where each time $R_0$ is each galaxy's own 
stellar optical disk scale length), while the dashed and dash-dotted 
curves show the velocities that the Newtonian and the linear potentials 
would separately produce. The dotted curves show the total velocities that 
would be produced without any adopted distance modification. No dark matter
is assumed.

\medskip
\noindent
Figure (2). The paths followed by waves in a COW type interferometer.

\medskip
\noindent
Figure (3). The paths followed by waves in a double slit experiment.

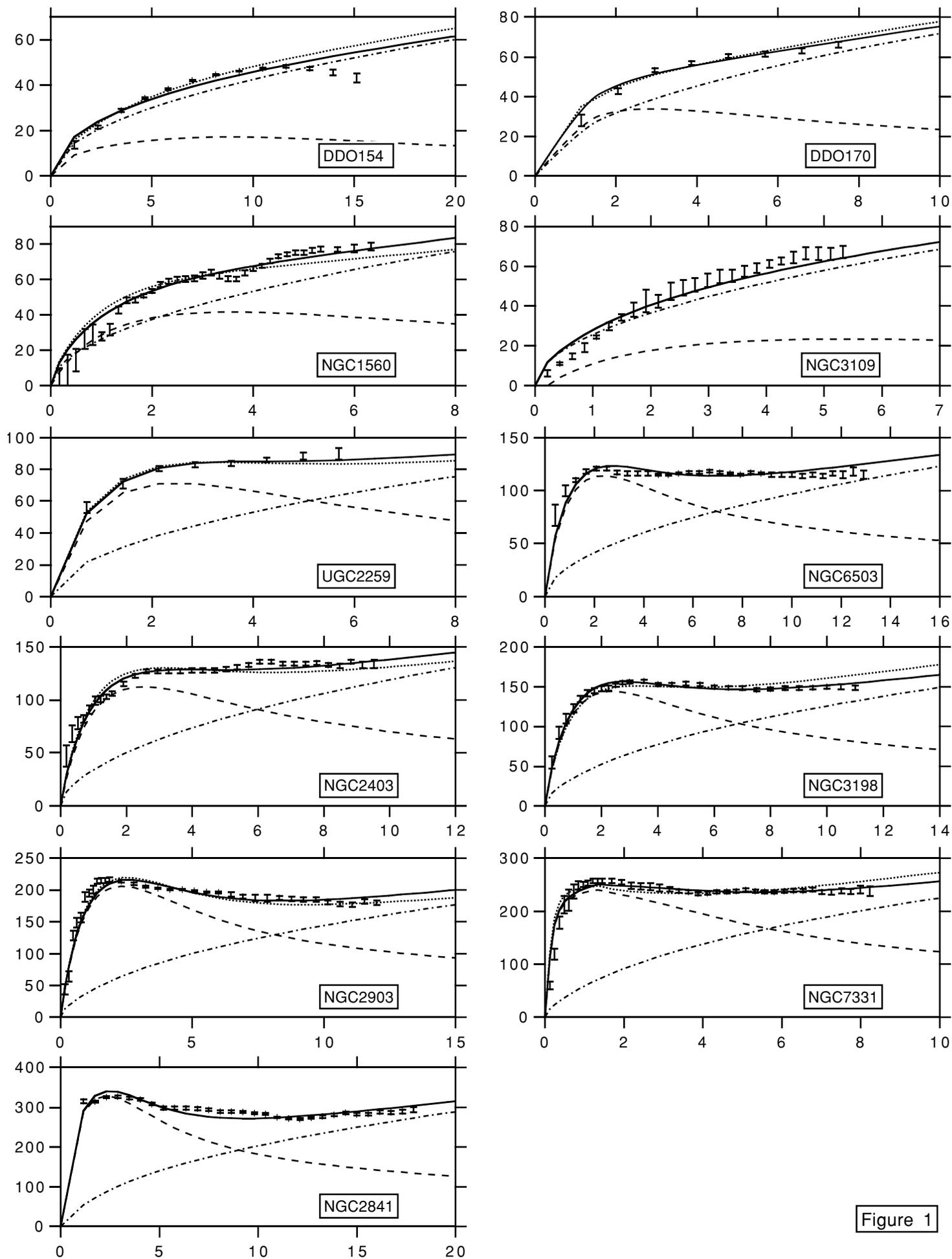
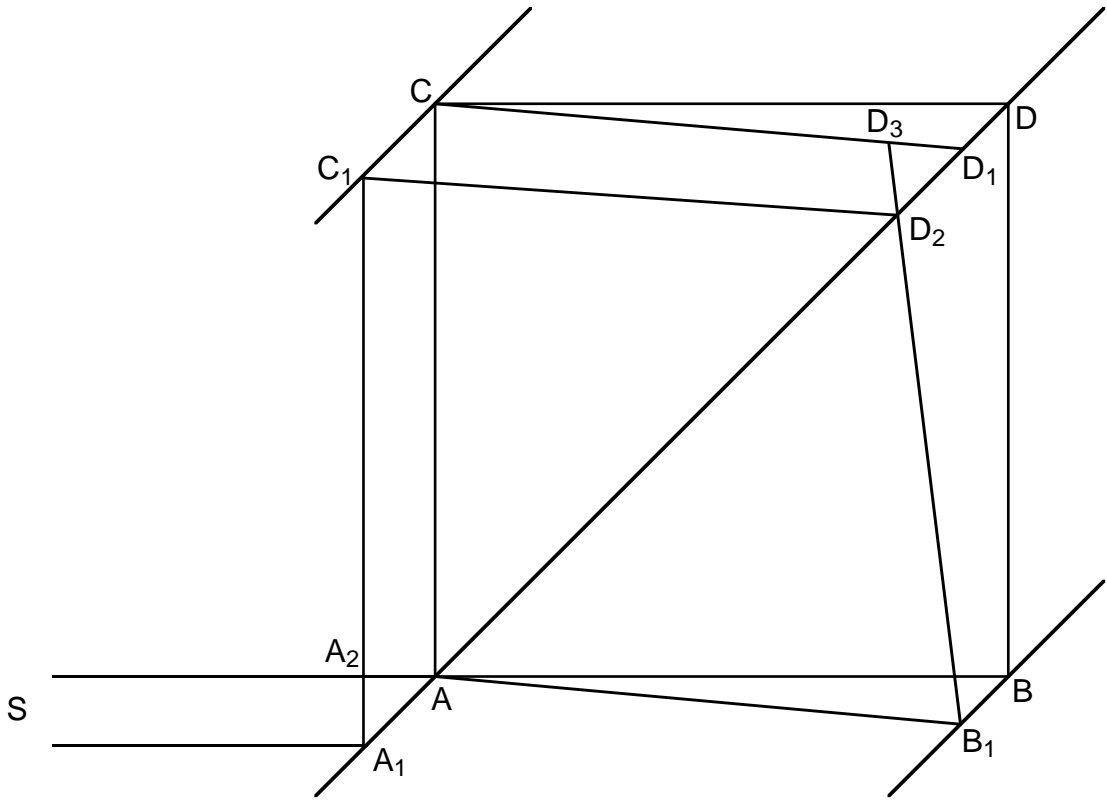
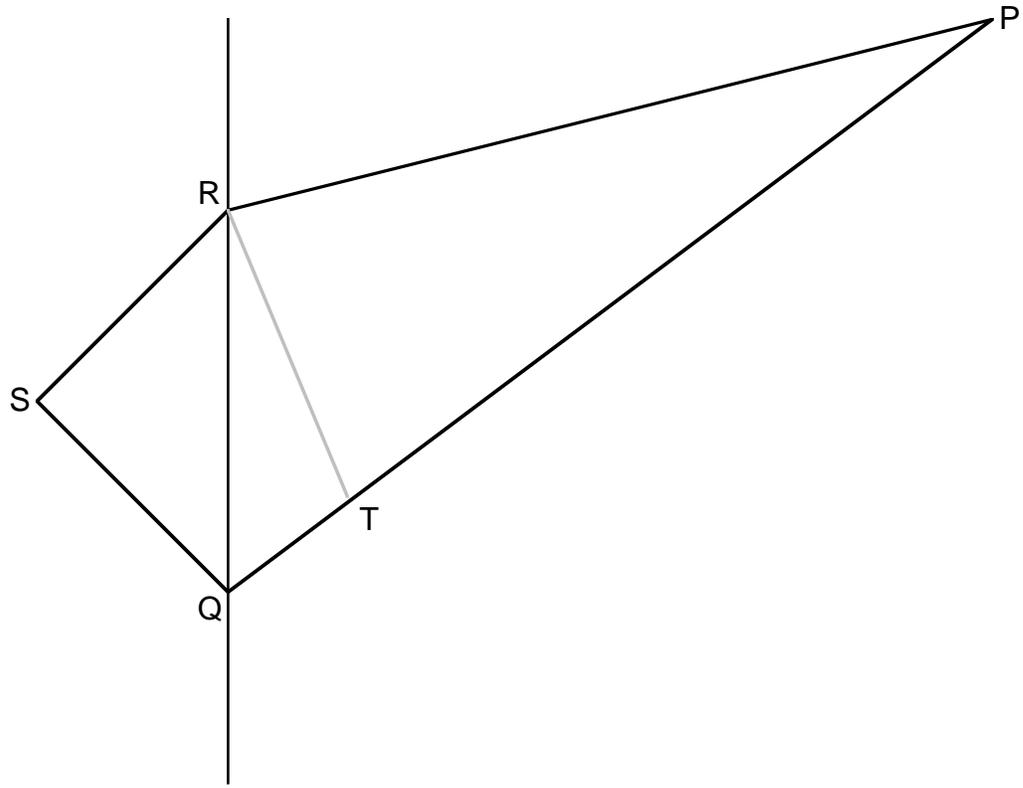
\end{document}